\documentclass[aps,prb,twocolumn,superscriptaddress]{revtex4-2}
\usepackage{graphicx}
\usepackage{float}
\usepackage{amsmath,amsfonts}
\usepackage{xcolor}
\usepackage{bbold}
\usepackage[colorlinks=true,allcolors=blue]{hyperref}
\usepackage{ulem}
\usepackage{braket}

\begin{document}
\title{Spin-momentum locking breakdown on plasmonic metasurfaces}

\author{Fernando Lor\'en}
\email{loren@unizar.es}
\affiliation{Instituto de Nanociencia y Materiales de Arag\'on (INMA), CSIC-Universidad de Zaragoza, 50009 Zaragoza, Spain\looseness=-1}
\affiliation{Departamento de F\'isica de la Materia Condensada, Universidad de Zaragoza, 50009 Zaragoza, Spain\looseness=-1}

\author{Cyriaque Genet} 
\affiliation{University of Strasbourg and CNRS, CESQ \& ISIS (UMR 7006), 8, all\'ee G. Monge, 67000 Strasbourg, France}

\author{Luis Mart\'in-Moreno}
\email{lmm@unizar.es} 
\affiliation{Instituto de Nanociencia y Materiales de Arag\'on (INMA), CSIC-Universidad de Zaragoza, 50009 Zaragoza, Spain\looseness=-1}
\affiliation{Departamento de F\'isica de la Materia Condensada, Universidad de Zaragoza, 50009 Zaragoza, Spain\looseness=-1}

\begin{abstract}
We present a scattering formalism to analyze the spin-momentum locking in structured holey plasmonic metasurfaces. It is valid for any unit cell for arbitrary position and orientation of the holes. The spin-momentum locking emergence is found to originate from the unit cell configuration. Additionally, we find that there are several breakdown terms spoiling the perfect spin-momentum locking polarization. We prove that this breakdown also appears in systems with global symmetries of translation and rotation of the whole lattice, like the Kagome lattice. Finally, we present the excitation of surface plasmon polaritons as the paramount example of the spin-momentum locking breakdown.
\end{abstract}

\maketitle

\section{Introduction}
Metasurfaces based on plasmonic arrays have been demonstrated to have a plethora of applications \cite{chen2016areview, genevet2017recent} such as sensing \cite{beruete2019terahertz}, imaging \cite{watts2014terahertz, walter2017ultrathin}, or telecommunications \cite{zhang2015gbps}. In particular, geometric phase metasurfaces (GPMs) have gained significant attention in the last years due to their ability to manipulate the polarization of light waves in a controllable manner \cite{zhao2011manipulating, yu2012abroadband, chervy2018room, fox2022generalized, singh2022topological}. One important property of these metasurfaces is that they can exhibit spin-momentum locking (SML), which refers to the coupling between the polarization and the momentum of the involved light waves \cite{bliokh2015spinorbit}. 

Despite the evidenced applicability of these plasmonic GPMs and numerous numerical studies, no first principles rigorous theoretical analysis had been developed. There have been studies for continuously space-variant structures \cite{bomzon2002spacevariant} and for structures with translation and rotation symmetries of the whole lattice under stringent conditions for the direction of the electric field \cite{shitrit2013spinoptical}. Recently, we have applied a scattering formalism to study holey plasmonic GPMs that present a chiral arrangement in the unit cell \cite{loren2023microscopic}.

This article presents a general analysis of the SML on GPMs, extending our previous study to lattices that present full translation and rotation symmetry. In particular, we apply it to the Kagome lattice, which has been considered as a platform for GPMs \cite{shitrit2013spinoptical, proctor2021higherorder} and also studied due to its relevance in antiferromagnets \cite{harris1992possible, schweika2007approaching}. Our results provide a comprehensive understanding of the SML mechanism on GPMs and have important implications for designing and optimizing these metasurfaces. Based on this general formalism, we demonstrate that the appearance of the SML breakdown is ubiquitous for any system, revealing the interplay between the SML and the linear character of the surface plasmon polaritons (SPPs). The SML breakdown appears in systems with and without global rotation symmetries, both of which will be considered below. 


\section{Theoretical formalism}
The general derivation of the scattering formalism used in this paper is provided in the Supplemental Material of \cite{loren2023microscopic}. In this section, we present the essential elements required to comprehend the relevant terms of the formalism, along with the article's results.

We consider a general plasmonic metasurface, this is, a metal slab characterized by a periodically repeated unit cell with an arbitrary number of elements ($N$) distributed in. A huge variety of shapes can be considered \cite{lmm2001theory, koerkamp2004strong, gordon2004strong, garciavidal2010light} yet we will focus on one of the simplest ones, rectangular dimples, which corresponds to the study of our metasurfaces by reflection. Analyzing them by transmission, if we had considered holes, would lead to the same main results. Each dimple has a short side $a$, a long side $b$, and depth $d$. Furthermore, each dimple is defined by its position ($\vec{r}_{\alpha} = (x_{\alpha}, y_{\alpha})^T$) and the angle with respect to the $\vec{u}_x$ direction ($\theta_{\alpha}$), where $\alpha$ is the index associated with each dimple.

\begin{figure}[ht]
\includegraphics[width=0.8\columnwidth]{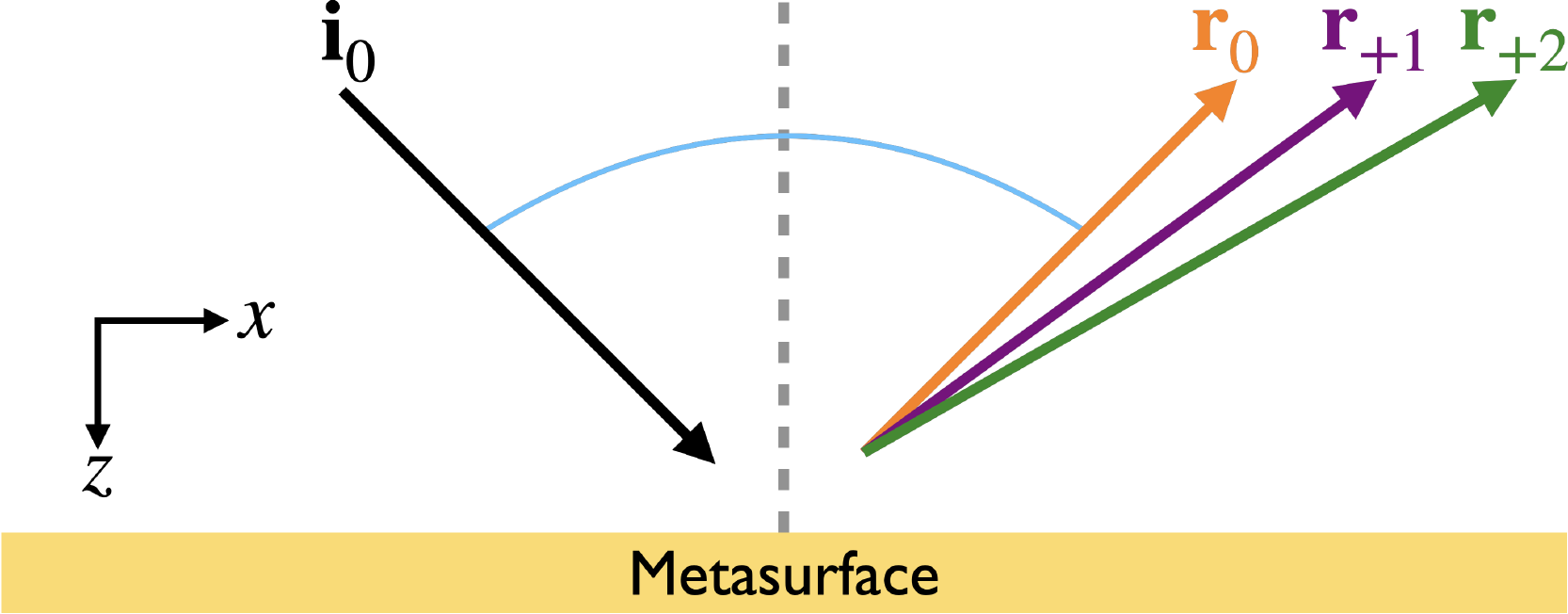}
\caption{Scheme of the excitation of the metasurfaces. $\mathbf{i}_0$ is the amplitude of the incident plane wave and $\mathbf{r}_m$ are the reflection coefficients of the Bragg modes.} 
\label{fig:Scheme}
\end{figure}

An electromagnetic (EM) plane wave is impinging our metasurface with an in-plane wavevector $\vec{k}^{in} = k^{in}_x \, \vec{u}_x + k^{in}_y \, \vec{u}_y$ and an incident polarization $\sigma_{in}$, and our goal is to compute the reflection coefficients into the different Bragg orders (see Figure~\ref{fig:Scheme}). For this purpose, we employ the coupled-mode method (CMM), which has been extensively used in the study of EM properties in metallic dimple arrays \cite{lmm2008minimal, garciavidal2010light, loren2023microscopic}. The CMM expands the EM fields in plane waves in the free space regions and waveguide modes inside the dimples, and finds the electric field amplitudes by properly matching the EM fields at the interfaces. 

The reciprocal lattice vectors that define our unit cell in the Fourier space are $\vec{G}_1$ and $\vec{G}_2$. The Bragg modes are characterized by an in-plane wavevector $\vec{k}_m = \vec{k}^{in} + m_1 \, \vec{G}_1 + m_2 \, \vec{G}_2$ and a polarization $\sigma$. We will combine the integers $m_1$ and $m_2$ into a single index: $m = (m_1, m_2)$, for notational simplicity.

We describe the behavior of the metal using the Perfect Electric Conductor (PEC) approximation, which assumes that the metal's dielectric constant
tends toward negative infinity. This simplification allows for a clearer description of the system's physics, as demonstrated in \cite{lmm2008minimal, loren2023microscopic}.
The effects of field penetration in the metal and associated losses are discussed in Appendices~\ref{app:theo}, \ref{app:SIBC}, and \ref{app:approx}, showing that the results obtained within the PEC approximation are qualitatively accurate.

It is convenient to express the polarization of each Bragg mode on the circular polarization (CP) basis to study the SML provided by our metasurface. We represent the reflection coefficients as spinors to contain both spin components: $\mathbf{r}_{m} = (r_m^+, r_m^-)^T$, where $\pm$ denote the right- and left-handed polarization (or spin), each of them defined within the plane perpendicular to the wavevector associated to the Bragg mode $m$. This representation is chosen because the spin of a plane wave is conserved upon reflection by a mirror \cite{bliokh2013dual, cameron2012optical, cameron2014optical, cameron2017chirality} (while the helicity changes sign).

The reflection coefficients in the CP basis with respect to the propagation directions satisfy the following equations
\begin{equation}
\mathbf{r}_m = - \delta_{m0} \, \mathbf{i}_0 + C_{m0} \,  Y_0 \, \mathbf{i}_0 - \sum_{m'} C_{mm'} \, Y_{m'} \, \mathbf{r}_{m'}.
\label{eq:rm}
\end{equation}
The first term is the specular reflection, being $\mathbf{i}_0$ the amplitude of the incident plane wave and $\delta_{m0}$ the Kronecker delta. $C_{mm'}$ are the \textit{geometric couplings} \cite{loren2023microscopic}, which are $2 \times 2$ matrices operating in polarization space. They couple different Bragg modes ($m'$ with $m$) via scattering with the plasmonic metasurface and encode the geometry of the unit cell through the overlaps between the Bragg and the waveguide modes. 

$Y_{m'}$ are also $2 \times 2$ matrices representing the modal admittances. They relate the in-plane magnetic field to the electric one and, in the CP basis, can be written as $Y_{m'} = \bar{Y}_{m'} \, \mathbb{1} + \Delta_{m'} \, \sigma_x$, where $\mathbb{1}$ and $\sigma_x$ are the $2 \times 2$ unit matrix and the Pauli matrix that swaps spin states, respectively. In terms of the linear p (transverse magnetic) - s (transverse electric) polarized basis, $\bar{Y}_{m'} \equiv (Y_{m' p} + Y_{m' s})/2$ and $\Delta_{m'} \equiv (Y_{m' p} - Y_{m' s})/2$. For a plane wave with frequency $\omega$ and in-plane wavevector $k_{m'} = |\vec{k}_{m'}|$ propagating in a uniform medium with dielectric constant $\epsilon$, the modal admittances are $Y_{m'p} = \epsilon / q_{m'z}$ and $Y_{m's} = q_{m'z}$, where $q_{m'z}= \sqrt{\epsilon - q_{m'}^2}$ ($q_{m'} = c \, k_{m'} / \omega$ and $c$ is the speed of light). Notice that $\Delta_0 = 0$ at normal incidence, while both $\bar{Y}_{m'}$ and $\Delta_{m'}$ diverge at the Rayleigh points (i.e., whenever a diffractive order becomes tangent to the metal-dielectric interface).

The geometric couplings allow us to explore the SML emergence because they provide the coupling between two different Bragg modes and their corresponding CP components. They can be written as
\begin{equation}
C_{mm'} = R^{k(m) \leftarrow z} C_{mm'}^z R^{z \leftarrow k(m')}.
\end{equation}
The interaction of the Bragg modes is ruled through the dimples, so the in-plane EM fields are the ones playing a role in the couplings. Therefore, the origin of the SML resides in the properties of the geometric couplings in the CP basis but with respect to the $\vec{u}_z$ direction, $C_{mm'}^z$. However, each Bragg mode is transversal so its polarization is defined with respect to the propagation direction. Then, we need the $R$'s to encapsulate the change of basis with respect to the $\vec{u}_z$ and the propagation direction. 

The matrix that changes basis from the $\vec{u}_z$ direction to the propagation direction of the $m$-th Bragg mode is $R^{k(m)\leftarrow z} = \frac{1}{2} \left[ (\sqrt{q_{mz}^2 + q_m^2}/q_{mz} + 1) \, \mathbb{1} + (\sqrt{q_{mz}^2 + q_m^2}/q_{mz} - 1) \, \sigma_x \right]$. The presence of $\sigma_x$ in $R^{k(m)\leftarrow z}$ implies its occurrence in $C_{mm'}$, leading to the swapping of spin states.

On the other hand, the expression for $C_{mm'}^z$ is:
\begin{equation}
	C_{m m'}^{z} = C' \sum_{\alpha=0}^{N-1} S_{m\alpha} S_{m' \alpha}^*,
	\label{eq:geometric}
\end{equation}
where $C'$ is the dimple cross-section, which depends on the dimple area and depth, and the impedance of the waveguide mode; and $S_{m\alpha}$ is a geometrical factor that measures how well a given EM plane wave overlaps with the fundamental mode in the dimple (details in Appendix~\ref{app:theo} and \cite{loren2023microscopic}).

Both $\sigma_x$ appearances (in $Y_{m'}$ and $C_{mm'}$) contribute to the mixing of the spin components of the Bragg modes, reducing the SML contrast and producing what we coined as \textit{spin-momentum locking breakdown} in \cite{loren2023microscopic}. We have shown that the SML breakdown terms are ubiquitous to any configuration independent of whether they host or not global rotation symmetries. The paramount example of the relevance of the SML breakdown is the excitation of SPPs because both $\Delta_{m'}$, and the factor $(\sqrt{q_{mz}^2 + q_m^2}/q_{mz} - 1)$) appearing in $R$, rise and become as large as $\bar{Y}_{m'}$ and $(\sqrt{q_{mz}^2 + q_m^2}/q_{mz} + 1)$).

In the succeeding sections, we will describe two different, although related, structures: without and with global rotation symmetries. For both, we will present the SML mechanism derived from their geometric couplings and the SML breakdown effects.


\section{Spatially rotated dimples along $\vec{u}_x$ direction}
\begin{figure*}[ht]
\includegraphics[width=\textwidth]{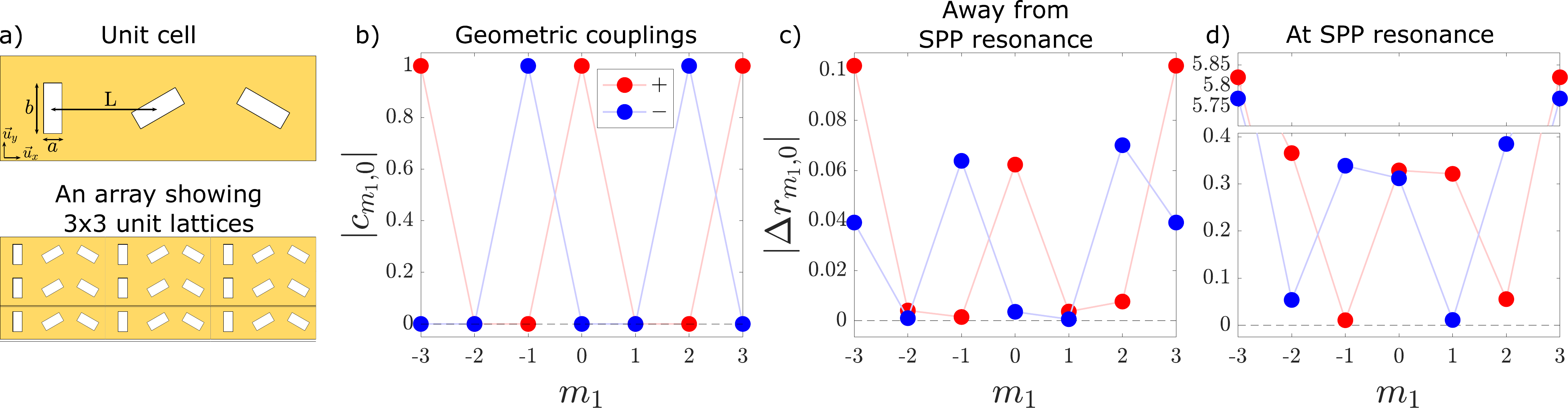}
\caption{(a) Unit cell with $N=3$ spatially rotated dimples and $n_w = 1$, and scheme of the considered full array, showing $3\times3$ unit lattices. (b) Geometric couplings with respect to the $\vec{u}_z$ direction, considering $m'_1 = m'_2 = m_2 = 0$ and spinor $+$. The spin $+/-$ component is represented in red/blue. (c, d) $|\Delta \mathbf{r}_m|$ with respect to $m = m_1$ and $m_2 = 0$, computed by taking a normal incident plane wave with spin $+$. Chosen geometrical parameters: $L = 460 \, nm$, $a = 80 \, nm$, $b = 220 \, nm$, and $d = 60 \, nm$. (c) is computed away from an SPP resonance, with incoming energy $\omega = 3 \, eV$. (d) is for an SPP resonance associated to the Bragg modes $m_1 = \pm 3$, with incoming energy $\omega = 2.692 \, eV$. We break the y-axis for a better observation of every Bragg mode.
} 
\label{fig:Chain}
\end{figure*}

We consider a rectangular unit cell of $N=3$ dimples evenly spaced along the $\vec{u}_x$ of the unit cell, with $L$ being the distance between the centers of the two nearest dimples, in both $x$- and $y$- directions. We consider that $\theta_{\alpha}$ varies linearly with $\alpha$: $\theta_{\alpha} = 2 \pi n_w \alpha/ N$, where the winding number $n_w$ defines the number of complete $2\pi$ rotations along the unit cell. We have selected the following set of geometrical parameters for the remainder of the paper: $L = 460 \, nm$, $a = 80 \, nm$, $b = 220 \, nm$, and $d = 60 \, nm$. These particular parameters align with those employed in the experiments detailed in \cite{chervy2018room, loren2023microscopic}. However, we note that the specific dimensions and inter-distances of the dimples, although influencing the dimple cross-section, do not impact the system’s topological properties.

The system is depicted in Figure~\ref{fig:Chain}a, where the winding number is $n_w = 1$. The case presented in \cite{loren2023microscopic} is similar and the appearance of SML breakdown was already demonstrated. The choice of $N=3$ and $n_w = 1$ is based on the system considered in the next section, the Kagome lattice, whose unit cell can be seen as three clusters of three dimples each, with winding numbers of $n_w=1$ as well. Another reason for considering $N=3$ is because the rotation steps of $2\pi/3$ are very far from the adiabatic and continuous condition required to apply the Berry phase formalism, which was conceived to analyze adiabatic and continuous deformations of a closed spatial path \cite{berry1984quantal}.

Notice that although the dimples perform a step-wise rotation along the unit cell, the whole lattice does not support global rotation symmetry. 

For this case, the reciprocal lattice vectors are: $\vec{G}_1 = 2\pi / (N L ) \, \vec{u}_x$ and $\vec{G}_2 = 2\pi / L \,\vec{u}_y$. Considering $m_2  = m _2' = 0$ is enough to explore the underlying physics because there is no inversion symmetry breaking along the $\vec{u}_y$ direction \cite{shitrit2013spinoptical, loren2023microscopic}. Thus, we consider $m = m_1$, $m' = m_1'$ and $k_y^m = k_y^{m'} = 0$. Besides, the small-dimple approximation simplifies the overlapping integrals by considering the dimples much smaller than the wavelength. Then, $C_{mm'}^z$ reads
\begin{equation}
\begin{aligned}
C_{mm'}^z &= C \sum_{\alpha = 0 }^{2} e^{i 2 \pi \alpha (m' - m) / N} 
\begin{pmatrix}
1 & e^{-i 2 \pi 2 n_w \alpha / N}\\
e^{i 2 \pi 2 n_w \alpha / N} & 1 \\
\end{pmatrix}\\
& = C N \left(\delta_{m, m' + n_0 N} \, \mathbb{1} + \sum_{s = \pm} \delta_{m,m' + n_0 N - 2n_w s} \,  \sigma_s \right),
\end{aligned}
\label{eq:chain3}
\end{equation}
where $n_0$ is any integer, $\sigma_{\pm} $ are Pauli matrices that increase and decrease spin, respectively, and $C = 4 a  b C' / (\pi ^2 A_{uc})$, being $A_{uc}$ the area of the unit cell. 

The SML mechanism is derived exactly from Eq.~\ref{eq:chain3}. The first term corresponds to the spin-preserving processes and the associated Bragg law is $k_x^{out} = k_x^{in} + n_0 \, G^0$, with $G^0 = 2\pi/L$. Two Bragg modes with a difference in indices proportional to N can be coupled if the spin is preserved. The second term describes the spin-flipping processes and the associated Bragg law is $k_x^{out} = k_x^{in} + n_0 G^0 \mp k_g$, where $k_g = 2\pi 2n_w / (NL)$ is the geometric momentum. Two Bragg modes with a difference in indices proportional to $N \pm 2 \, n_w$ can be coupled if the spin is changed to $\mp 1$, which is exactly the spin-to-momentum conversion of the SML.

To illustrate this, we come with spin $+ \equiv (1,0)^T$ and represent both spin components of the normalized amplitudes of the geometric couplings in the CP basis. This is, $\mathbf{c}_{m_1,0} \equiv (c_{m_1, 0}^+,c_{m_1, 0}^-)^T = C^z_{m 0} \cdot (1,0)^T / (CN)$.

In Figure~\ref{fig:Chain}b, we represent $|c_{m_1,0}^{\pm}|$. The SML is evident. Spin is preserved for $m_1 = 0, \pm 3$, which are multiples of $N$; and spin is flipped for $m_1 = 2,-1$, which are $2 \, n_w$ and $2 \, n_w - N$ respectively. Hence, the exact SML mechanism arises from the geometric couplings with respect to the $\vec{u}_z$ direction, $C_{mm'}^z$. 

When computing the full EM system (reflection coefficients), breakdown terms appear in both geometric couplings and modal admittances. Additionally, there is the contribution from the specular reflection. As we want to study the interaction of the light with the dimple lattice, we define $\Delta \mathbf{r}_m = \mathbf{r}_m + \delta_{m0} \, \mathbf{i}_0$, which removes the specular reflection from the zero order for a better observation of the SML breakdown.

In Figure~\ref{fig:Chain}c we represent $|\Delta r^{\pm}_{m_1,0}|$ for an incoming plane wave impinging normally to the metasurface with spin $+$ and energy $\omega = 3 \, eV$. The consequences of the SML breakdown terms are already noticeable: all the Bragg modes are a combination of both CP states, and the perfect SML does not hold anymore but is recognizable. Since at that frequency SPP resonances are not excited, the general behavior is still similar to the perfect SML.


Note that we are studying a plasmonic metasurface and the breakdown is maximum when a plasmonic resonance is excited \cite{loren2023microscopic}. Thus, we show the reflection coefficients when we are at a plasmonic resonance in Figure~\ref{fig:Chain}d. We represent $|\Delta r^{\pm}_{m_1,0}|$ at a SPP resonance associated to the Bragg modes $m_1 = \pm 3$. We use an incoming plane wave impinging normally to the metasurface with spin $+$ and energy $\omega = 2.692 \, eV$. The consequences of the SML breakdown terms are now predominant: $|\Delta r^{\pm}_{\pm3,0}|$ are very large and both spin components are similar, which is characteristic of the linearly $p$ polarized character of the SPP. Moreover, the perfect SML behavior cannot be recognized because of SML breakdown, being spoiled and mixed both spin components of all the Bragg modes.


\section{Kagome lattice}

\begin{figure*}[ht]
\includegraphics[width=\textwidth]{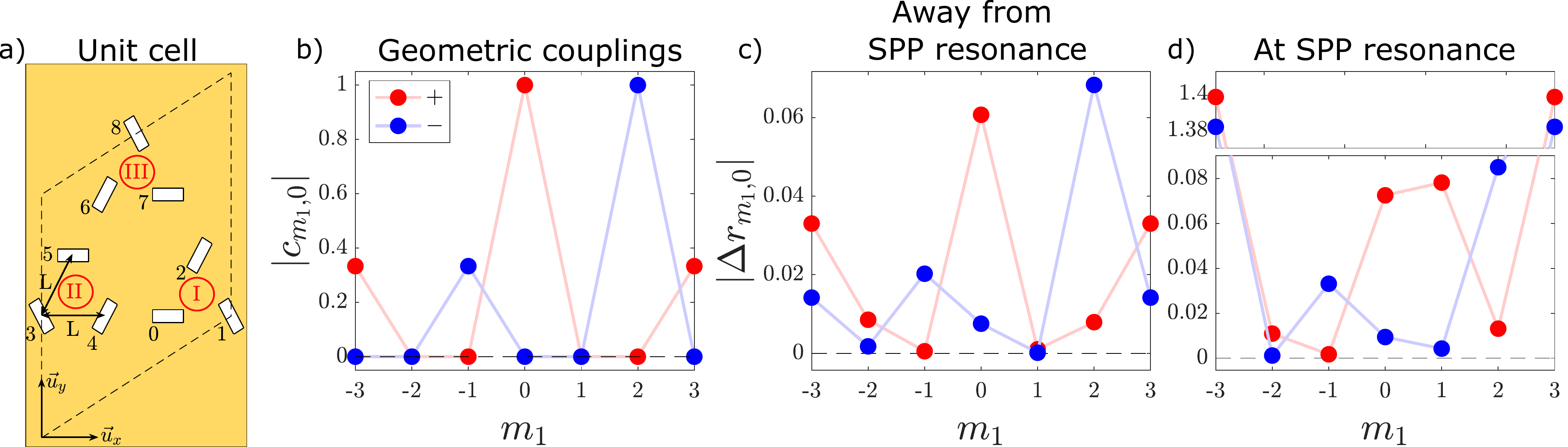}
\caption{(a) Unit cell of the $\sqrt{3} \times \sqrt{3}$ Kagome lattice, defined by the dashed line with the three clusters highlighted in red. (b) Geometric couplings with respect to the $\vec{u}_z$ direction, considering $m'_1 = m'_2 = m_2 = 0$ and spinor $+$. The spin $+/-$ component is represented in red/blue. (c, d) $|\Delta \mathbf{r}_m|$ with respect to $m = m_1$ and $m_2 = 0$. Computed by taking a normal incident plane wave with spin $+$. Chosen values: $L = 460 \, nm$, $a = 80 \, nm$, $b = 220 \, nm$, and $d = 60 \, nm$. (c) Computed away from an SPP resonance, with incoming energy $\omega = 3 \, eV$. (d) Computed at an SPP resonance associated with the Bragg modes $m_1 = \pm 3$, with incoming energy $\omega = 2.694 \, eV$. We break the y-axis for a better observation of every Bragg mode.} 
\label{fig:Kagome}
\end{figure*}

In this section, we present the main result of the article: the appearance of the SML breakdown in a system with combined translation and rotation symmetry of the whole lattice. This is the staggered (or $\sqrt{3}\times \sqrt{3}$) Kagome lattice (KL) \cite{grohol2005spin, schweika2007approaching, shitrit2013spinoptical}. The reciprocal lattice vectors of the KL are: $\vec{G}_1 = 2 \pi / (3L) \, \vec{u}_x$ and $\vec{G}_2 = \pi / (3L) (- \vec{u}_x + \sqrt{3} \vec{u}_y)$. We will analyze its geometric couplings, as well as the reflection coefficients.


This symmetry is important because it has been used in other works \cite{shitrit2013spinoptical} to study the appearance of SML via group theory arguments, although restricted to waves with an electric field perpendicular to the surface and at normal incidence. 

Figure~\ref{fig:Kagome}a shows a schematic representation of the considered KL. The unit cell is defined by the dashed lines and is composed of $N=9$ dimples, defined by the positions of their centers and their angles with respect to the $\vec{u}_x$ direction (see Table~\ref{tab:Kagome} in Appendix~\ref{app:table}). These nine dimples can be subdivided in three similar clusters $\{\alpha\} = \{\{0,1,2\},\{3,4,5\},\{6,7,8\}\}$. The dimples in each cluster are distributed forming an equilateral triangle, with angles that are step-wisely rotated with a winding number of $n_w = 1$. 

Each triangular cluster has the same number of dimples and the same winding number as the rectangular unit cell of the previous section. However, they have different spatial distributions. Consequently, the involved Bragg modes in the KL host similar, but different, coupling processes.

The geometric couplings in the CP basis with respect to the $\vec{u}_z$ in the PEC and small-dimple approximations are,

\begin{widetext}
\begin{equation}
\begin{aligned}
	C_{mm'}^{z} = & \, C \sum_{\alpha=0}^{8} e^{i (\vec{k}_{m'} - \vec{k}_{m}) \vec{r}_{\alpha}}
	 \begin{pmatrix}
	c_{++} & c_{+-} \, e^{- i 2 \theta_{\alpha}} \\
	c_{-+} \, e^{ i 2 \theta_{\alpha}} & c_{--} \\
	\end{pmatrix}\\
	= & \, C \, A^{mm'}
	\begin{pmatrix} 
 	 c_{++} \, \delta_{m_1 + m_2, m_1' + m_2' + 3 \, n_0} &  - c_{+-} \, \delta_{m_1 + m_2, m_1' + m_2' + 3 \, n_0 - 2 \, n_w}\\
	-c_{-+} \, \delta_{m_1 + m_2, m_1' + m_2' + 3 \, n_0 + 2 \, n_w} &  c_{--} \, \delta_{m_1 + m_2, m_1' + m_2' + 3 \, n_0}\\
	\end{pmatrix},
		\label{eq:kagomeSML}
\end{aligned}
\end{equation}
\end{widetext}

where we have defined $c_{\sigma \sigma' } = (\vec{k}_m \cdot \vec{\sigma}) (\vec{\sigma}' \cdot \vec{k}_{m'}) / (k_m k_{m'})$, being $\vec{\sigma} = \vec{u}_x + i \sigma \vec{u}_y$, with $\sigma =  \pm$. These $c_{\sigma \sigma'}$ are the projections of the Bragg modes $m$ and $m'$ with the circular polarizations $\sigma$ and $\sigma'$, respectively. The Kronecker deltas provide the selection rules between these Bragg modes, being $n_0$ an integer. Besides, depending on the Bragg modes to be coupled, the coupling amplitude is different: $|A^{mm'}| = N$ if both $\Delta_{1}$ and $\Delta_2$ are even, and $|A^{mm'}| = N/3$ in the rest of cases; being $\Delta_{1/2} = m_{1/2}'-m_{1/2}$. This is inferred from the sum over the dimples in the unit cell, in the first line of Equation~\ref{eq:kagomeSML}.

Equation~\ref{eq:kagomeSML} rules two different processes. One process (given by the diagonal elements of $C^z_{mm'}$) conserves spin. The corresponding Bragg law, called \textit{standard} Bragg law \cite{shitrit2013spinoptical} is $\vec{k}^{out} = \vec{k}^{in} + m_1 \, \vec{G}_1 + m_2 \, \vec{G}_2$ such that $m_1 + m_2  = 3 \, n_0$ (notice that the incident plane wave corresponds to $m_1' = m_2' = 0$) . The other process flips spin (off-diagonal elements of $C^z_{mm'}$). The corresponding Bragg law, called \textit{spin-orbit} Bragg law \cite{shitrit2013spinoptical}, satisfies another condition: $m_1 + m_2 = 3 \, n_0 \mp 2 \,  n_w$, which is exactly the SML mechanism.

Figure~\ref{fig:Kagome}b shows the SML mechanism derived from the geometric couplings. We represent $\mathbf{c}_{m_1,m_2} = C^z_{m 0} \cdot (1,0)^T / (CN)$, where $(1,0)^T$ is the spinor for the spin $+$. Although we have considered both $m_1$ and $m_2$ in the calculation, we take $m_2 = 0$ for a simpler representation. We observe the feature of the coupling amplitudes $A^{mm'}$ of the different processes. It is easy to observe that the SML mechanism that we described above is satisfied.

Once we have shown how the SML arises from the geometric couplings for the KL, we look at $\Delta \mathbf{r}_m$. In Figure~\ref{fig:Kagome}c we represent $|\Delta r^{\pm}_{m_1,0}|$ for an incoming plane wave with spin $+$, energy $\omega = 3 \, eV$ and normal to the metasurface. Since SPPs are not excited at that frequency, the general behavior is similar to the perfect SML, although we already see some signatures of the breakdown. The amplitude relation between the different modes is no longer exactly satisfied, and we also observe small amplitudes of modes that should be zero if SML were exact. 

Finally, in Figure~\ref{fig:Kagome}d, we show the reflection coefficients when a plasmonic resonance is excited. We represent $|\Delta r^{\pm}_{m_1,0}|$ at a SPP resonance associated to the Bragg modes $m_1 = \pm 3$. We use an incoming plane wave with spin $+$, energy $\omega = 2.694 \, eV$ and impinging normally to the metasurface. The SML breakdown terms have acquired a governing relevance. $|\Delta r^{\pm}_{\pm3,0}|$ are very large and both spin components are similar, which is characteristic of the linearly $p$ polarized character of the SPP. From these resonantly excited modes, successive couplings with other modes can occur. In consequence, we cannot recognize anymore the expected SML because both spin components of all the Bragg modes are spoiled and mixed. 

The physical interpretation is as follows: the EM fields carry CP light perpendicular to the propagation direction of the plane waves. However, the system has a particular symmetry perpendicular to the planar metasurface ($\vec{u}_z$ direction). This mismatching results in that when the CP light gets projected onto the planar surface, it becomes elliptical (which is a combination of the two CP states) and then, the SML is spoiled.


\section{Conclusion}
We have shown that even a system with combined translation and rotation symmetry of the whole lattice suffers spin-momentum locking breakdown. The physical interpretation lies in the elliptical projection onto the planar metasurface of the circularly polarized light. Therefore, together with the results obtained in \cite{loren2023microscopic}, this shows that any system, with or without global lattice symmetries, presents breakdown of the SML. Nonetheless, we stress that the breakdown terms are often small, so the SML is a useful concept. However, in some cases such as the plasmonic resonances, breakdown terms become very relevant. Plasmon resonances are, thus, the paramount example of SML breakdown.

Despite the occurrence of this breakdown, it presents an opportunity to optimize the system in order to minimize it. Additionally, other applicative perspectives could be renewed by the consideration of the results presented in this work, such as optovalleytronic systems \cite{li2021experimental}, non-linear hybrid metasurfaces \cite{hu2019coherent}, and topology-based high-resolution sensors \cite{ding2017gradient}.

\section*{Acknowledgements}
We acknowledge Project PID2020-115221GB-C41 was financed by MCIN/AEI/10.13039/501100011033 and the Aragon Government through Project Q-MAD.

This work is part of the Interdisciplinary Thematic Institute QMat of the University of Strasbourg, CNRS, and Inserm. It was supported by the following programs: IdEx Unistra (ANR-10-IDEX-0002), SFRI STRATUS project (ANR-20-SFRI-0012), and USIAS (ANR-10-IDEX-0002-02), under the framework of the French Investments for the Future Program.


\appendix

\section{Details of the theoretical formalism} \label{app:theo}
Here, we extend the calculations presented in the main text and introduce the required quantities such as $C'$ and the overlapping integrals.

We present the formalism within the surface impedance boundary conditions (SIBC) approximation. The SIBC approximation provides a more accurate derivation because it considers the real dielectric constant of the metal $\epsilon_M (\omega)$, via the Lorentz-Drude model \cite{vial2005improved}, and also the penetration of the EM fields into the metal through the surface impedance $z_s = 1/ \sqrt{\epsilon_M}$. Yet, we consider $z_s = 1/ \sqrt{\epsilon_M + 1}$, which is a phenomenological correction that leads to the exact dispersion relation of surface plasmon polaritons (SPPs) in a metal-vacuum interface. The reflection coefficients are now

\begin{equation}
f^+_m \, \mathbf{r}_m = - f_0^- \,\delta_{m0} \, \mathbf{i}_0 + C_{m0} \, Y_0 \, \mathbf{i}_0 - \sum_{m'} C_{mm'} \, Y_{m'} \, \mathbf{r}_{m'},
\label{eq:rm_SIBC}
\end{equation}
where the SIBC signatures are encapsulated in the geometric couplings and in the quantities $f_m^{\pm}$, which are $2\times2$ matrices in the CP basis with respect to the propagation of the $m$-th Bragg mode, that depend on the surface impedance $z_s$ such that:
\begin{equation}
	f_m^{\pm} = \frac{1}{2} \begin{pmatrix}
	f_{mp}^{\pm}+f_{ms}^{\pm} & f_{mp}^{\pm}-f_{ms}^{\pm}\\
	f_{mp}^{\pm}-f_{ms}^{\pm} & f_{mp}^{\pm}+f_{ms}^{\pm}\\
	\end{pmatrix},
\end{equation}
with $f_{m\sigma}^{\pm} = 1 \pm z_s Y_{m\sigma}$ and $\sigma = \{ p,s\}$.
 
The dependence of the metal approximation in the geometric couplings is encapsulated in the constant $C'$:
\begin{equation}
	C'_{SIBC} = \frac{1}{Y} \frac{f^+ \, f^- \, (1+\Phi)}{f^+ - f^- \, \Phi} ,
\end{equation}
whereas
\begin{equation}
	C'_{PEC} = \frac{1}{Y} \frac{1+\Phi}{1-\Phi},
\end{equation}
being $Y$ the modal admittance of the fundamental waveguide mode, $f^{\pm} = 1 \pm z_s \, Y$, $\Phi = - e^{i2 k_z^w d}$ and $k_{z}^{w}$ is the propagation constant along the z-direction of the fundamental waveguide mode.  For a rectangular dimple with long side $b$, filled with a material with dielectric constant $\epsilon_{d}$,  $k_z^w = \sqrt{\epsilon_{d} (\omega/c)^{2} - k_w^2}$,  with $k_w = \pi / b$.

We posed in the main text that the geometric couplings depend on the overlapping integrals $S_{m\sigma\alpha}$ between the Bragg modes (characterized by $m$ and $\sigma$) and waveguide modes (characterized by the dimple index $\alpha$). A general expression for the overlapping integrals is intricate because of the dependence on the in-plane momenta and the size of the dimples (it can be found in \cite{loren2023microscopic}). However, if we consider the small-dimple approximation for which the dimple size is smaller than the wavelength, they read

\begin{equation}
	S_{m \sigma \alpha} = \sqrt{\frac{a b}{2 A_{uc}}} \frac{4}{\pi} \, v_{m \sigma \alpha} \, e^{-i \vec{k}_m \vec{r}_{\alpha}},
	\label{eq:overlaps-small}
\end{equation}
where $A_{uc}$ is the area of the unit cell, $\sigma$ is the polarization of the considered Bragg mode, and $v_{m p \alpha} = (k_x^m \cos{\theta_{\alpha}} + k_y^m \sin{\theta_{\alpha}}) / k_m$ and $v_{m s \alpha} = (-k_y^m \cos{\theta_{\alpha}} + k_x^m \sin{\theta_{\alpha}}) / k_m$, being $k_x^m$ and $k_y^m$ the $x$ and $y$ components of the in-plane momentum $\vec{k}_m$, respectively.

With these expressions, one can easily achieve the geometric couplings for both systems presented in the main text (see Equations~\ref{eq:chain3} and \ref{eq:kagomeSML}).


\section{Kagome lattice elements} \label{app:table}
In Tab.~\ref{tab:Kagome}, we present the defining quantities for all the dimples comprising the analyzed Kagome lattice. We label each dimple with an index $\alpha$ and show its center position and its angle.

\begin{table}[ht]
\begin{tabular}{c|c|c|c|c}
$\alpha$ & $x_{\alpha}$ & $y_{\alpha}$ & $\vec{r}_{\alpha}$ & $\theta_{\alpha}$ \\ \hline
$0$ & $2L$ & $\sqrt{3} L$ & $2\vec{R}_1 /3 + \vec{R}_2/6$ & $\pi/2$ \\
$1$ & $3L$ & $\sqrt{3} L$ & $\vec{R}_1$ & $7 \pi /6$ \\
$2$ & $5L/2$ & $3\sqrt{3} L/2$ & $5\vec{R}_1 /6 + \vec{R}_2/3$ & $11\pi/6$ \\
$3$ & $0$ & $\sqrt{3} L$ & $\vec{R}_1 /2 $ & $7\pi/6$ \\
$4$ & $L$ & $\sqrt{3} L$ & $\vec{R}_1 /3 + \vec{R}_2/3$ & $11\pi/6$ \\
$5$ & $L/2$ & $3 \sqrt{3} L/2$ & $\vec{R}_1 /6 + 2\vec{R}_2/3$ & $\pi/2$ \\
$6$ & $L$ & $2\sqrt{3} L$ & $\vec{R}_1 /3 + 5\vec{R}_2/6$ & $11\pi/6$ \\
$7$ & $2L$ & $2\sqrt{3} L$ & $2\vec{R}_1 /3 + 2\vec{R}_2/3$ & $\pi/2$ \\
$8$ & $3L/2$ & $5\sqrt{3} L/2$ & $\vec{R}_1 /2 + \vec{R}_2$ & $7\pi/6$
\end{tabular}
\caption{Centers positions (also in terms of the direct lattice vectors $\vec{R}_1 = 3 L \vec{u}_x + \sqrt{3} L \vec{u}_y$ and $\vec{R}_2 = 2 \sqrt{3} L \vec{u}_y$) and angles for the $N=9$ dimples constituting the unit cell of the $\sqrt{3} \times \sqrt{3}$ KL represented in Figure~\ref{fig:Kagome}a.}
\label{tab:Kagome}
\end{table}


\section{SIBC approximation in the KL} \label{app:SIBC}

\begin{figure}[ht]
  \includegraphics[width=\columnwidth]{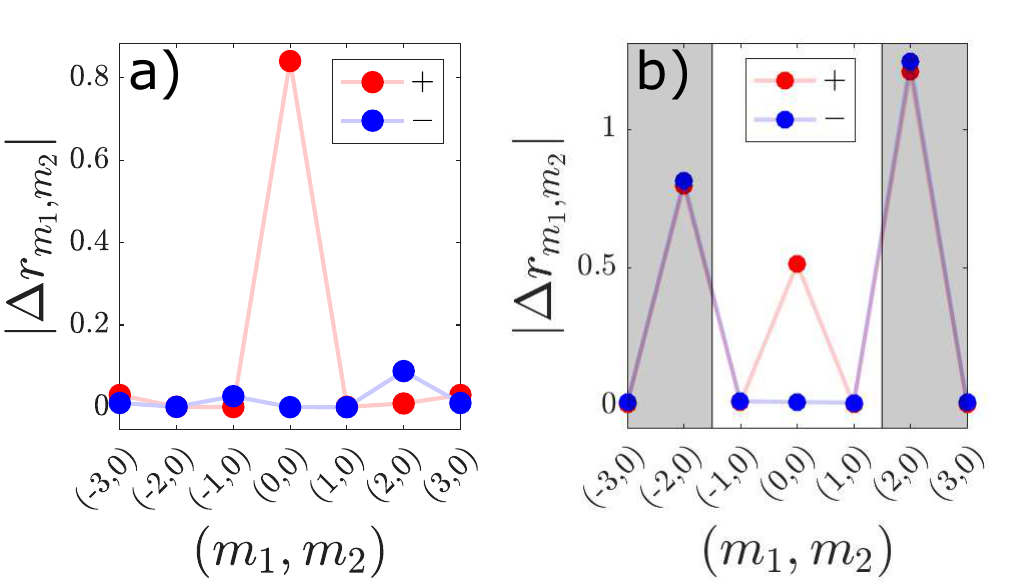}
  \caption{$\Delta \mathbf{r}_m$ with respect to $m = (m_1,m_2)$. The spin $+/-$ component is represented in red/blue, and we have considered $m_2 = 0$. It has been computed taking a normal incident plane wave with spin $+$. Chosen geometrical parameters: $L = 460 \, nm$, $a = 80 \, nm$ $b = 220 \, nm$, and $d = 60 \, nm$. We have considered the SIBC approximation, phenomenologically enlarging the dimple dimensions by 1.25 times the skin depth to consider the EM field penetration in the metal \cite{lmm2001theory}. a) Away from any plasmonic resonance, with $\omega = 3 \, eV$ energy for the incident plane wave. b) At the plasmonic resonance associated with the Bragg modes $m_1 = \pm 2$, with $\omega = 1.73 \, eV$ energy for the incident plane wave. The shadowed region indicates modes that are outside the light cone.}
  \label{fig:SIBC_hierarchyM20}
\end{figure}

In this section, we expand on the SML breakdown cases that we studied in the main text for the Kagome lattice. We compute the effect of considering the SIBC approximation and finite-size dimples. This is shown in Figure~\ref{fig:SIBC_hierarchyM20}, where we consider a representative case of non-resonant excitation and another case of a resonant plasmonic excitation. In both cases, we represent $\Delta r_m$ and sweep $m_1$. We observe the effect of the SIBC approximation at first glance. The zero order is larger than the rest (except when we excite an SPP and the resonant modes govern). Furthermore, the SML breakdown is evident in both figures, although the underlying SML can be noticed in the orders $m_1 = -1,2$ of Figure~\ref{fig:SIBC_hierarchyM20}a where the spin $-$ component is larger than the spin $+$ one, for instance. Besides, in Figure~\ref{fig:SIBC_hierarchyM20}b we observe the same behavior of very large $|\Delta r_{\pm2}^{\pm}|$ as we presented in the main text for $|\Delta r_{\pm3}^{\pm}|$. Therefore, in the SIBC approximation, the SML becomes less evident because of the metal absorption.

In this case and below, we have kept $m_2 = 0$ not only for the representation but also for the simulation. This does not affect the physical behavior because the $\vec{G}_1$ direction presents a breaking of the inversion symmetry \cite{shitrit2013spinoptical, loren2023microscopic}.


\section{Analysis of the incident momentum in the KL}
 \begin{figure}[ht]
    \includegraphics[width=0.8\columnwidth]{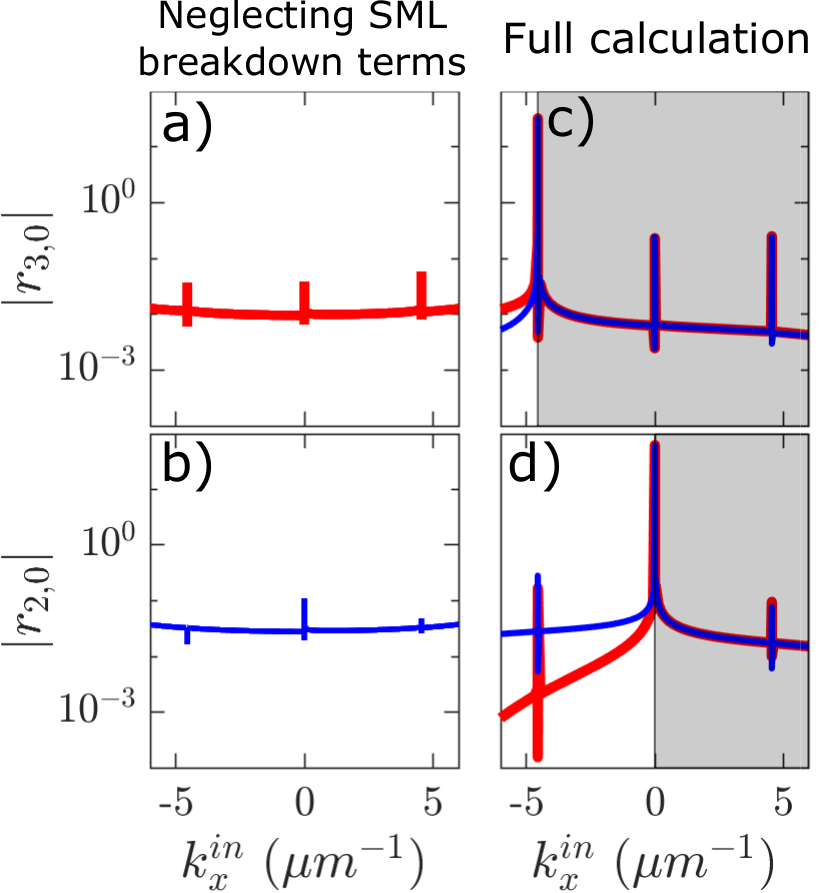}
    \caption{Absolute value for both spin components (red/blue is spin $+/-$) for two reflection coefficients $r_{3,0}$ and $r_{2,0}$. We take $m_2 = 0$, $k_y^{in} = 0$, $\omega = 1.79 \, eV$ and incoming spin $+$. (a, b) are computed neglecting SML breakdown terms. (c, d) are computed considering the SML breakdown terms (full calculation). We take the PEC and the small-dimple approximations. Chosen geometrical parameters: $L = 460 \, nm$, $a = 80 \, nm$ $b = 220 \, nm$, and $d = 60 \, nm$. The shadowed region indicates that reflection coefficients are outside the light cone for that incident momentum $k_x^{in}$.}
    \label{fig:refcoeff_M20}
\end{figure}

We have focused on the KL by analyzing its SML, the breakdown terms, and its dependence on being or not at a plasmonic resonance. For the latter analysis, we have varied the energy and kept the normal incidence. However, we can also excite different SPP resonances by varying the incident momentum. This section will show how the reflection coefficients behave when the incident momentum is varied away from the normal.

Figure~\ref{fig:refcoeff_M20} represents the absolute value of both spin components for two reflection coefficients: $r_{3,0}$ and $r_{2,0}$, with respect to the incident momentum in the $x$ direction: $k_x^{in}$. We have chosen the representative values of $k_y^{in} = 0$, $\omega = 1.79 \, eV$ and $\sigma_{in} = +$. Figures~\ref{fig:refcoeff_M20}a and~\ref{fig:refcoeff_M20}b show that there is only one spin component for each mode, which is in excellent agreement with the SML features derived from the geometric couplings $C^{z}_{mm'}$. The three small peaks for each subfigure correspond to plasmonic resonances which, given that breakdown terms have been neglected, preserve the SML. However, when we perform the full calculation, considering all SML breakdown terms, both spin components are non-negligible and the SML is spoiled (see Figures~\ref{fig:refcoeff_M20}c and~\ref{fig:refcoeff_M20}d). Besides, when the corresponding plasmonic resonance is associated with the Bragg mode that we are representing via the reflection coefficient, there is an enhancement of the latter. This was also seen in the $|\Delta \mathbf{r}_m|$ plots of the main text.

Logically, the SML breaks down when a plasmonic resonance is excited because the SPPs are linearly $p$ polarized. However, this breakdown persists even when $k_x^{in}$ is increased away from resonance. The reason is that for larger $k_x^{in}$, the Bragg modes associated with these reflection coefficients ($r_{3,0}$ and $r_{2,0}$) are evanescent. Given this and considering that both breakdown sources (modal admittances and the change of basis matrices) depend on the momentum in the $\vec{u}_z$ direction of the corresponding Bragg mode $q_{mz}$, it is easy to infer that the evanescent modes introduce a strong breakdown as well.

\section{Analysis of the approximations in the KL} \label{app:approx}
\begin{figure}[ht]
    \includegraphics[width=0.8\columnwidth]{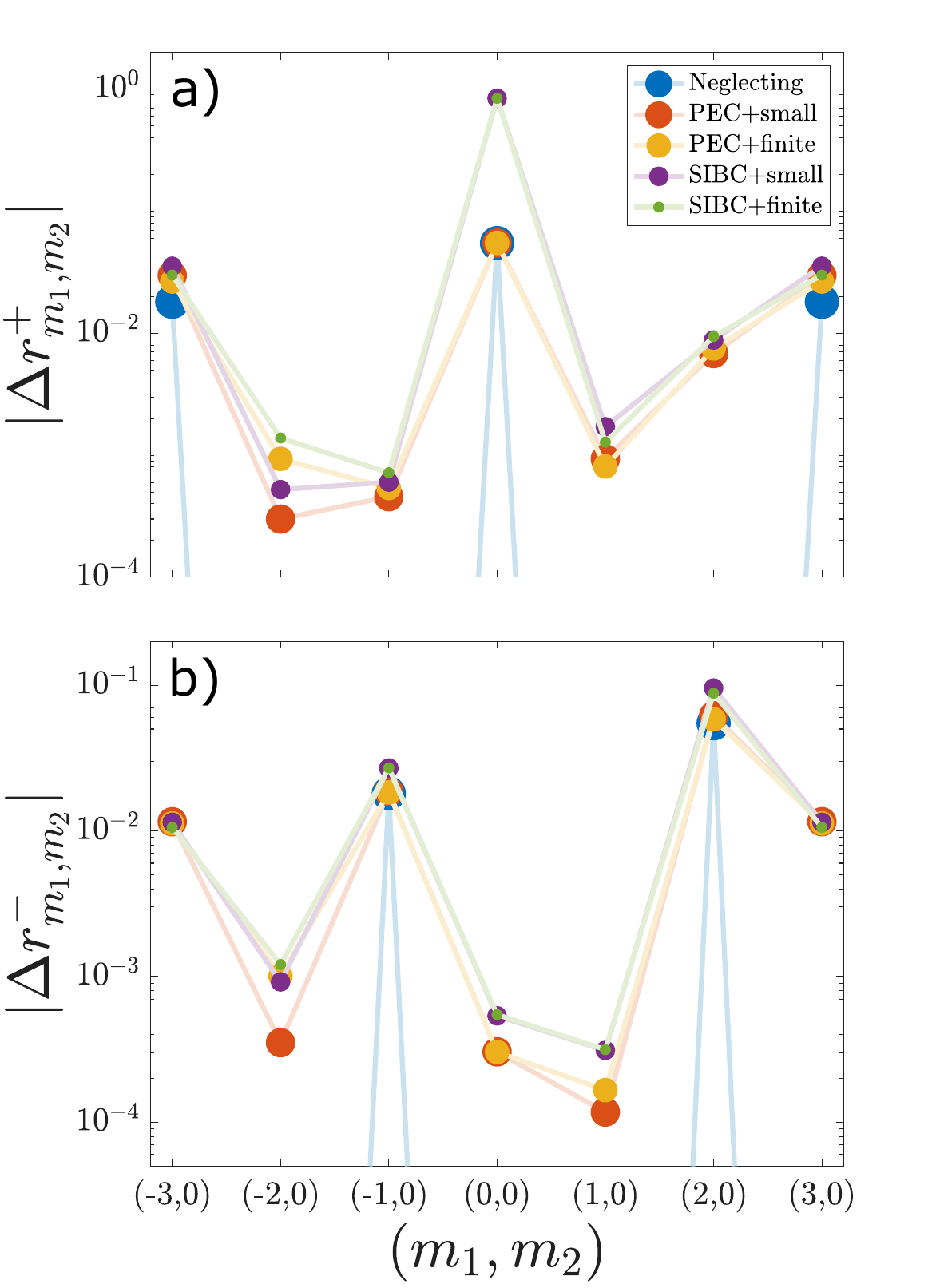}
    \caption{Both spin components of $\Delta \mathbf{r}_m$. a) $\Delta \mathbf{r}_m^+$, spin $+$ component. b) $\Delta \mathbf{r}_m^-$, spin $-$ component. They have been computed taking a normal incident plane wave with spin $+$ and energy $\omega = 3 \, eV$,  and we have considered $m_2 = 0$. Chosen geometrical parameters: $L = 460 \, nm$, $a = 80 \, nm$ $b = 220 \, nm$, and $d = 60 \, nm$.}
    \label{fig:comparison_approx}
\end{figure}

The results presented in the main text are computed in the PEC and small-dimple approximations. On the other hand, in Figure~\ref{fig:SIBC_hierarchyM20} we showed what happens if we calculate the same quantities but in the SIBC approximation and with finite-size dimples. A global comparison is still lacking. For this reason, in Figure~\ref{fig:comparison_approx}, we display the five possibilities: neglecting the SML breakdown terms (blue), PEC and small-dimple (red), PEC and finite-size (yellow), SIBC and small-dimple (purple), and SIBC and finite-size (green).

Along the main text and the rest of the appendices, we have dealt with two of the five approximations detailed in Figure~\ref{fig:comparison_approx}. In Figures~\ref{fig:Chain}c,~\ref{fig:Chain}d,~\ref{fig:Kagome}c,~\ref{fig:Kagome}d,~\ref{fig:refcoeff_M20}c and~\ref{fig:refcoeff_M20}d, we considered the PEC and small-dimple approximations, or what we call ``full calculation". Besides, in Figure~\ref{fig:SIBC_hierarchyM20}, we used the SIBC and finite-size approximations. Therefore, we present Figure~\ref{fig:comparison_approx} to compare them and add the rest of the possible combinations: neglecting SML breakdown terms, PEC with finite-size, and SIBC with small-dimple approximations. 

The effects of the different approximations are observed in Figure~\ref{fig:comparison_approx}, representing both spin $+/-$ components of the reflection coefficients. Blue dots represent the case of neglecting SML breakdown terms; because of that, some modes are zero (not seen). This approximation is equivalent to the behavior of the geometric couplings $C^z_{mm'}$. The rest of the approximations represent different levels of SML breakdown. The smallest SML breakdown is obtained when the metal is considered as a PEC and the dimples are very small, whereas the maximal breakdown appears when the metal is real and the dimples are finite-sized. Moreover, a general pattern appears: the effect of the dimple size is less relevant than the effect of the PEC approximation. That is to say, choosing small dimples or finite dimples only provides a small deviation over the reflection coefficients. However, a greater difference appears between the PEC and the SIBC approximations.

Note that we have stayed away from any plasmonic resonance for this comparison because the plasmonic resonance locations depend on the considered metal approximations.
 
\bibliographystyle{apsrev4-2}
\bibliography{draft}

\end{document}